\documentclass[acmsmall,screen,nonacm]{acmart}
\authorsaddresses{}

\AtBeginDocument{%
  \providecommand\BibTeX{{%
    \normalfont B\kern-0.5em{\scshape i\kern-0.25em b}\kern-0.8em\TeX}}}

\usepackage{booktabs}
\usepackage{ragged2e}
\usepackage{multirow}
\usepackage{enumitem}

\begin{document}

\title[Minimum Viable Ethics]{Minimum Viable Ethics: From Institutionalizing Industry AI Governance to Product Impact}

\author{Archana Ahlawat}
\email{archana.ahlawat@princeton.edu}
\affiliation{%
  \institution{Princeton University Center for Information Technology Policy}
  \country{USA}
  \city{Princeton}
}

\author{Amy Winecoff}
\email{aw0934@princeton.edu}
\affiliation{%
 \institution{Center for Democracy and Technology}
  \country{USA}
  \city{Washington, D.C.}
}

\author{Jonathan Mayer}
\email{jonathan.mayer@princeton.edu}
\affiliation{%
  \institution{Princeton University Center for Information Technology Policy}
  \country{USA}
  \city{Princeton}
}

\begin{abstract}

Across the technology industry, many companies have expressed their commitments to AI ethics and created dedicated roles responsible for translating high-level ethics principles into product. Yet it is unclear how effective this has been in leading to meaningful product changes. Through semi-structured interviews with 26 professionals working on AI ethics in industry, we uncover challenges and strategies of institutionalizing ethics work along with translation into product impact. We ultimately find that AI ethics professionals are highly agile and opportunistic, as they attempt to create standardized and reusable processes and tools in a corporate environment in which they have little traditional power. In negotiations with product teams, they face challenges rooted in their lack of authority and ownership over product, but can push forward ethics work by leveraging narratives of regulatory response and ethics as product quality assurance. However, this strategy leaves us with a \textit{minimum viable ethics}, a narrowly scoped industry AI ethics that is limited in its capacity to address normative issues separate from compliance or product quality. Potential future regulation may help bridge this gap. 
\end{abstract}

\ccsdesc[500]{Human-centered computing~Empirical studies in HCI}

\keywords{AI ethics, organizational change, product impact, industry practice}

\maketitle
\section{Introduction}

In recent years, a nascent class of AI ethics and governance professionals have joined technology companies to work on identifying and mitigating risks from AI. At the same time, AI policy has gained steam with several prominent proposals for regulation worldwide~\cite{execorder,eu_ai_act}. As regulatory standards develop alongside frontier AI systems, major developers of AI products and systems influence direction both for policy and what its implementation in practice looks like. Emerging regulatory approaches that rely on internal compliance will necessarily depend on ethics professionals' judgment and abilities to wield power internally. The resourcing of new ethics and governance roles alongside coinciding AI principles may signal that companies are motivated to address values and societal impacts of AI carefully~\cite{microsoft_rai_standard, google_rai_princip}. But some critics view the trend as ethics-washing, suggesting that these do not lead to significant impact on technology development or business decisions~\cite{metcalf_boyd_owning_ethics,tech_ethics_layoffs}. Still, in the absence of concrete regulations or a robust third-party auditing ecosystem, corporate ethics and governance work is clearly a critical first-line component of building systems that are safe, fair, and aligned with societal values.

AI ethics and governance professionals attempt to reshape the values embedded in company decisions through new practices. As the field develops, it is unclear whether these initiatives are truly leading to robust organizational change or a state of ad-hoc ethics. In order to shed light on this, we investigate the challenges and strategies these professionals face in institutionalizing practices. We then interrogate how effectively they translate these attempts at organizational change into product impact. Throughout the paper, we refer to AI ethics and governance simply as ``AI ethics'' as shorthand since it reflects the way many professionals tackling these issues are branded. As global discussion on these topics has increased, ``AI governance'' is increasingly used as a catch-all for the types of activities described in this paper, but is not yet standard. 

As prior literature has shown, AI ethics teams and professionals face challenging organizational dynamics, limited industry standards, and little clarity on how to bring theoretical ideas of responsibility, fairness, and safety to an applied context~\cite{Rakova_2021,madaio2022assessing}. This “principles to practices gap” is difficult to overcome given the contested and emergent conceptual and technical standards of ethics in AI as well as a lack of proper organizational investment~\cite{schiff2020principles}. Additionally, the mandate and level of decision-making power of AI ethics teams are precarious. Recent interview-based studies on AI ethics organizational practices uncover challenges ethics professionals face in persuading product teams to prioritize mitigations and the informal strategies they use to shore up support~\cite{Ali_2023,Rakova_2021}. 

By viewing the development of industry AI ethics through the lens of institutional change literature, we can conceptualize AI ethics professionals as institutional entrepreneurs, or ethics entrepreneurs as per Ali et. al.~\cite{Rakova_2021,Ali_2023}. They work within the constraints of current institutional structures in attempt to infuse them with new values. Adopting this analytical lens, interview-based research focused on organizational change in the technology industry in the context of privacy, security, and UX practices similarly exposes the challenges and strategies at play. Analogous data privacy literature is especially useful, since this work shows how organizational structures and practices of privacy professionals are often symbolic or ritualistic. They achieve superficial compliance with regulation but fail to meaningfully protect consumers~\cite{ari_ground,ari_lawreview}. 

This study extends prior work by interrogating the challenges and strategies of AI ethics professionals as they aim to institutionalize generalizable practices, from organizational structures to technical tools and frameworks. Given that AI ethics initiatives are more mature than a few years ago, we can now better analyze this process. Ultimately, we evaluate whether and how industry AI ethics practices, from developing to institutionalized, robustly and consistently impact products. To this end, we ask the following research questions:

 \begin{itemize}
     \item \textbf{RQ1:} How effective are AI ethics professionals at institutionalizing their organizational and technical practices?
     \begin{itemize}
         \item \textbf{RQ1a:} What challenges do AI ethics professionals encounter as they aim for institutionalization?
         \item \textbf{RQ1b:} What strategies do AI ethics professionals use as they aim for institutionalization?
     \end{itemize}
     \item \textbf{RQ2:} How effective are AI ethics professionals at influencing product decisions?
     \begin{itemize}
         \item \textbf{RQ2a: }What challenges do AI ethics professionals encounter in attempting to influence product decisions?
         \item \textbf{RQ2b:} What strategies do AI ethics professionals use to influence product decisions?
     \end{itemize}
 \end{itemize}

The goal of these questions is to draw a picture from practices to impacts on products. \textbf{RQ1} is focused on new practices that AI ethics professionals seek to put in place, while \textbf{RQ2} is centered on the outcomes of those practices. We decouple these in order to disentangle the practices themselves from how they manifest in product decisions. The goal of \textbf{RQ1} is to understand the process of institutionalization in industry AI ethics and how successful it is. The goal of \textbf{RQ2} is to understand the extent to which AI ethics professionals are able to change product development and deployment. The challenges and strategies revealed can inform practitioners of best practices and pitfalls. In order to answer these questions, we conducted semi-structured interviews with 26 U.S. based software engineers, product and project managers, mixed methods workers, and researchers who are currently working or have worked within the last three years on AI ethics at technology companies. Our study design is descriptive and inductive, as we are interested in the subjective experiences of individuals in these roles. 

Our interviews revealed a state of minimum viable ethics in industry, in which AI ethics professionals operate with a limited decision space. They are forced to prioritize interventions that conveniently slot into existing business or compliance narratives. They act opportunistically, using any and all tactics available to them, as they attempt to institute and scale up new technical and organizational practices. The challenges of this process include reliance on informal relationship-building, undefined engagement models with product teams, and gaps between AI ethics research and implementation. All these elements require AI ethics professionals to flexibly experiment with their methods. In order to institutionalize their practices, they relied on formalizing and normalizing product release processes, documentation, and reusable, comprehensible technical tools. 

Despite strategic and often successful efforts to scale, we find that there is still significant decoupling between practices and changes in product direction. This is due to lack of authority over product, difficulty communicating speculative harms, dependence on leadership support, and conflicts rooted in differing ethics and product team goals, from features to deployment timelines. AI ethics professionals can lean on narratives of regulatory compliance and ethics as product quality assurance to motivate product changes. However, these narratives are insufficient for motivating action on a broader array of AI ethics concerns.

In this paper, we first discuss relevant prior literature on neo-institutional theory, organizational change as related to privacy, security, and UX practices, and finally, industry AI ethics (Section~\ref{sec:related_work}). We then describe the methods for our qualitative interview-based study (Section~\ref{sec:methods}). In the results (Section~\ref{sec:results}), we synthesize participant observations to uncover the challenges and strategies of AI ethics professionals as they aim to institutionalize their practices and to describe what institutionalization typically looks like. We then detail the challenges and strategies involved in translating AI ethics practices to product outcomes. Our results highlight limitations of these strategies in achieving impact. Turning to discussion (Section~\ref{sec:discussion}), we suggest lessons from similar challenges in the areas of privacy and UX and opportunities to improve practices. We close with an examination of policy implications for more effectively shifting industry priorities and with a sketch of areas for future scholarship.

\section{Related Work}
\label{sec:related_work}
Three areas of literature are particularly relevant to and motivation for this project. First, the broad scholarship on neo-institutional theory helps to conceptualize the role of individual agency and entrepreneurship in how organizations change. Second, specific to technology-related societal problems, the HCI and information security communities have examined how organizations address contested practices. This research illustrates how the lens of institutional entrepreneurship can apply to practitioners in technology companies. Third, and closest to this work, previous scholarship on AI ethics in organizational contexts highlights the need for ethics professionals and the challenges that they can encounter. We conclude the discussion of related work by directly linking our research questions to these areas of literature.

\subsection{Institutional change: legitimacy, decoupling, and institutional entrepreneurship}
Neo-institutional theory offers an interdisciplinary umbrella of explanatory frameworks that examine how formal and informal institutional structures interact with individual incentives and desires to shape how institutions develop. Particularly relevant to this study is the literature on how and why organizations change in response to economic, social, and governmental pressures~\cite{iron_cage,dimaggio_interest,embeddedagency}. Successfully adapting to satiate company stakeholders such as consumers, media, the mainstream public, and regulators legitimizes an organization~\cite{manage_legit,stakeholder_salience}. However, ``when adaptations to institutional pressures contradict internal efficiency needs, organizations sometimes claim to adapt when they in reality do not; they decouple action from structure in order to preserve organizational efficiency''~\cite{isomorphism_diffusion_decoupling}. Decoupling refers to a situation in which high-level policies and structures do not manifest directly in actions and outcomes. Externally, the existence of rhetorical commitments and official structures, such as in the case of AI ethics work, signals that an organization has adapted. Bureaucratic, siloed organizations and a lack of proper accounting of outcomes may even lead internal actors to believe this~\cite{institutionalized_organizations}. Yet, the actually existing ``ceremonial activity'' of organizations merely has ``ritual significance: it maintains appearances and validates an organization''~\cite{institutionalized_organizations}. This echoes discussions of ethics washing in the technology industry.

Although decoupling means that there is no institutionalized accountability for the translation of principles to outcomes, individual actors within companies may wield significant discretionary power to reshape their organizations~\cite{embeddedagency,orginst_chp}. Crozier's theory of bureaucracy highlights how institutions do not deterministically operate through formalized procedures and rules, but rather are contingent on individuals who operationalize these~\cite{crozier}. Similarly, the theory of street level bureaucracy, which refers to the loose coupling between policies and enforcement, reveals the critical role of first-line professionals who implement principles~\cite{street_level}. In an algorithmic context, Pääkkönen argues that discretionary decision-making is pivotal in points of uncertainty, ``where formal rules cannot be turned into actionable consequences without human judgment,'' for example when an engineer updates a ranking system in accordance with new content policies~\cite{bureaucracy_algo}. In these cases, practitioners decide how to interpret internal policy. 

Instead of a picture of static, top-down organizations, neo-institutional theory shows how institutions are reified through individuals' actions within defined but flexible institutional arrangements. This leaves significant room for institutional entrepreneurs to transform their organizations according to their values and interests~\cite{dimaggio_interest}. These internal entrepreneurs seek to ``dislodge existing practices, ... introduce new ones, and then ensure that these become widely adopted and taken for granted by other actors''~\cite{orginst_chp}. One key method of doing this is to ``[align] their work with the formal rules of the organization,'' such as by framing desired changes in the language of existing business priorities~\cite{crozier}. They further legitimize their initiatives by strategically forming coalitions across the organization, gaining resources to build them, and persuading stakeholders~\cite{embeddedagency}. Through this process, internal actors can reduce decoupling by enacting the symbolic changes in company values.

\subsection{Organizational change in privacy, security, and UX practices}
\label{sec:rel:othertechpractices}
Previous research in HCI and related fields has investigated other domains in which organizational practices in technology slowly evolved as a result of regulatory pressures and values-driven institutional entrepreneurship. Literature on privacy, security, and UX in industry detail how practitioners act as institutional entrepreneurs as they incorporate ethical considerations into the new job functions and practices associated with these domains. Such work shows ``how the interaction of structure and situated practices can unfold dynamics within an organization that eventually lead to sustainable change over time''~\cite{sec_routine}. As these fields developed, practitioners have tried to actively manage change every step of the way, seeking to transform ``unsystematic personal activity'' into standard company routines~\cite{sec_routine}.

In their attempts to influence the values and priorities of their companies, industry practitioners often take advantage of existing organizational narratives and structures. For instance, UX professionals who prioritize ethical design make their work legible to product owners by creating strong business cases to back them up, such as framing usability for underserved populations as an untapped market opportunity~\cite{softresistance_ux,ethicalmediation,designingrai_ux}. Similarly, privacy and security concerns are communicated as business risks or product quality issues~\cite{ari_ground,sec_routine,dims_ux,privrisks_investors}. Several studies describe new practices that served as values levers -- methods for introducing reflection and deliberation about values in design and technology teams~\cite{valueslevers}. These range from speculative red teaming activities to help product teams identify security vulnerabilities to ethics trainings~\cite{sec_routine,sec_fictions,sec_devworkshops}. To slowly embed new practices into teams, entrepreneurial ``vigilantes'' such as privacy or security champions became early adopters excited to test these out with teams~\cite{priv_champions,hodgepodgerai,ux_integration_casestudy}. Understanding how ``routines can become a source of change,'' security practitioners built code testing tools in hopes that more and more engineers would adopt them if they were easy to use and proved helpful~\cite{sec_routine}.

However, as this literature exposes, each step of introducing a new practice and progressively institutionalizing it is precarious and does not always end in long-term change~\cite{softresistance_ux}.  Without much institutional power, security, privacy, and UX practitioners may achieve some small-scale or one-off changes, but struggle to achieve broad buy-in to change how product teams make decisions. Recent work on the efficacy of privacy practices in particular expose how the supposed institutionalization of privacy obscures the reality of a fragmented and devalued privacy practice. Privacy professionals from front-line workers to Chief Privacy Officers innovated to create new privacy norms at their companies through internal audits and dedicated engineering teams~\cite{mulligan_bamberger}. Yet, Waldman contends that these workers are ``so constrained by antiprivacy organizational structures that their work ends up serving corporate surveillance interests in the end''~\cite{ari_ground}. In this analysis, technology companies fundamentally shape collective imagination on privacy. Furthermore, they systematically deprioritize privacy considerations in relation to monetizable product features by limiting privacy professionals' engagements with product. Symbols of compliance, such as extensive documentation, and an emphasis on procedures rather than specific required outputs result in weak privacy protection~\cite{ari_ground}. Studying how privacy practices have evolved in conversation with regulation, product teams, and privacy professionals provides a useful parallel to industry AI ethics.

\subsection{AI ethics in practice}
AI ethics as a field has matured, with groups across industry, academia, and the public sector iterating on technical developments, conceptual frameworks, and principles to help steer AI systems and products in a responsible manner~\cite{global_landscape_guidelines,whittlestone_principles,khan2021ethics,measurement_fairness}. A great deal of work in the AI ethics academic community focuses on technical components of ethics, such as documentation proposals, new methods, and tools. Examples are model cards and data sheets, methods for augmenting datasets to ensure equitable model performance, and interpretability tools to communicate why a particular decision was made~\cite{Mitchell_2019,gebru2021datasheets,interp_ref,10.1145/3543507.3583538}. On the conceptual side, frameworks for auditing sociotechnical systems abound~\cite{closing_acc_gap,nist_rmf}. In industry, several well-resourced technology companies have produced high-level AI principles and created AI ethics teams to operationalize these internally~\cite{microsoft_rai_standard,google_rai_princip}. Previous literature questions the usefulness of these abstract AI principles, which are not usually accompanied by concrete commitments, actions, or clear enforcement mechanisms~\cite{uselessness,Mittelstadt_2019}. This ``principles to practices'' gap is at the center of a growing area of research in AI ethics~\cite{schiff2020principles,audit_the_auditors}.

Workshop, survey, and interview-based studies in venues such as CSCW, CHI, FAccT, and AIES investigate this by studying ML practitioners at technology companies. This research uncovers the technical and organizational blockers they encounter, from difficulties translating academic research to real-world systems to divergent incentives of product teams and AI ethics teams~\cite{closing_acc_gap,Holstein_2019,madaio2022assessing,toolkits,andrus2021what,bakalar2021fairness}. For instance, on the technical front, responsibility tools may be used differently in practice from how they are intended to be used~\cite{userauditing_industrypractice,bhatt2020explainableml}. Studies on interpretability tools have shown that practitioners frequently misinterpret and overrely on them~\cite{datasci_interpretability}. As another example, ML practitioners sometimes retrofit their systems to use metrics that fairness toolkits provide even if they are not applicable to the systems at hand~\cite{fairnesstoolkits}. Organizational issues include lack of resourcing for interventions such as collecting better datasets for evaluations, struggles to incorporate end-user stakeholders, and persuading product owners and engineers to perform ethics work~\cite{madaio2022assessing,rismani2022plane,userauditing_industrypractice,metcalf_boyd_owning_ethics}. 

A growing body of work elaborates on these organizational difficulties. Data scientists and ML engineers formulate problems and execute projects through a constant cycle of negotiation amongst competing priorities and incentives~\cite{Passi_2019_problemformulation}. Rakova et al. identified organizational blockers and enablers for AI ethics work, highlighting weak and unaccountable decision-making structures paired with reactive and haphazard prioritization of projects as major blockers~\cite{Rakova_2021}. Building on this, recent work by Ali et al. points to the decoupling of AI ethics principles, practices, and structures from product outcomes. They characterize AI ethics professionals as ethics entrepreneurs who attempt to push organizational change~\cite{Ali_2023}. Several studies similarly emphasize the discretionary power of ML practitioners in defining what and how risks are prioritized~\cite{rismani2022plane,Veale_2018}. Deng et al. identified ``bridging'' and ``piggybacking'' strategies. In bridging, AI ethics teams use educational and documentation tools to communicate concepts like fairness to product teams. Piggybacking refers to co-opting existing institutional procedures to normalize ethics work, such as through checklists added to the traditional privacy review process~\cite{piggybacking}.

\subsection{Conceptual connections to this study's research questions}

In addressing \textbf{RQ1} in this study, we build on prior research by further investigating the institutionalization process for industry AI ethics practices. We view AI ethics professionals as values-driven institutional entrepreneurs who seek to fundamentally change the way product development and deployment decisions are made. As a low-ranking group of individuals aiming to impact large-scale, complex systems and processes, institutionalization is key to systemic, long-lasting change. One-off efforts are not enough to challenge existing ideologies and practices that structure technology companies~\cite{metcalf_boyd_owning_ethics}. We therefore foreground the institutional change framework. Parallels between AI ethics and privacy, security, and UX issues bolster this chosen analytical framework. Much like AI ethics, these fields are highly cross-functional and involve or involved emerging, unanswered normative, technical, and regulatory questions. As they have become more established, they have scaled up (e.g. high ranking positions like Chief Privacy Officers) and become standardized (e.g. security requirements). We see AI ethics developing through a similar process and aim to flesh this out further. In choosing this framework, we also echo recent work from Ali et al. on AI ethics entrepreneurship as well as Rakova et al.'s description of prevalent and aspirational practices of AI ethics professionals~\cite{Ali_2023,Rakova_2021}. Going beyond these, we systematically uncover how AI ethics may be transitioning from the current state to these aspirational practices. We explicitly detail the distinction between experimental, ad-hoc stages and their scaled up, standardized analogues.

In addressing \textbf{RQ2}, we analyze the challenges and strategies at play in translating new practices into outcomes: changes in product development. Inspired by previous work on the symbolic nature of institutionalized privacy practices, our analysis treats successful implementation of practices as distinct from their influence on product. We focus on product impact because products and services are the most direct ways that technology companies and their values affect society. Reflecting this, the day-to-day work of ``front-line'' AI ethics professionals, similar to privacy, security, and UX professionals, typically centers on specific products or systems. Although other elements such as marketing decisions and communications are important for shaping how AI impacts society, we view product as the primary and most dominant venue for translating ethics into practice. Putting this section in conversation with previous literature, we note that research often describes AI ethics teams in isolation from other teams or touches on product team relationships as one component of many organizational factors that impact AI ethics work. We narrow in on product-ethics team relationships to extend this work by detailing how and why, exactly, product teams can be blockers or enablers for AI ethics interventions.

By systematically uncovering challenges and strategies involved in the transition from ad-hoc practices to institutionalized practices, and then investigating the subsequent translation into product outcomes, we help develop a model of what effective institutionalization of AI ethics looks like. Adding to previous research, we use our analysis to offer recommendations and insights for research, organizational design, and practices to propel effective AI ethics work. 

\section{Methods}
\label{sec:methods}
\begin{table}
   \centering
  \caption{We interviewed participants in a diverse range of roles common on AI ethics teams -- engineering, PM, and research.}
  \label{tab:participant_roles}
    \renewcommand{\arraystretch}{1.5}
    \begin{tabular}{p{4cm} p{4.75cm} r}
        \toprule\\[-5.2ex]
                         Role& Participant IDs& Participants\\[-0.5ex]
        \midrule
          Software engineer, research engineer, ML engineer& P4, P7, P8, P11, P14, P18, P19, P25, P26& 9\\
 PM & P1, P3, P5, P6, P9, P12, P13, P17, P20, P23 &10\\
 Researcher & P2, P10, P15, P24 &4
    \end{tabular}
\end{table}
\begin{table}
   \centering
  \caption{The majority of participants were full-time employees, while a few contracted or volunteered to work on AI ethics issues.}
\label{tab:participant_employment_status}
  \renewcommand{\arraystretch}{1.5}
    \begin{tabular}{p{4cm} p{4.75cm} r}
        \toprule\\[-5.2ex]
                         Employment status & Participant IDs & Participants\\[-0.5ex]
        \midrule
          Full-time & P1, P2, P4, P5, P7, P8, P9, P10, P11, P12, P13, P14, P15, P16, P18, P19, P20, P22, P23, P24, P26 & 21\\
 Contract or volunteer work  & P3, P6, P17, P21, P25 &5\\
    \end{tabular}
\end{table}
\begin{table}
   \centering
\caption{The companies reflected in this study are primarily either enterprise or social media companies.}
\label{tab:company_types}
  \renewcommand{\arraystretch}{1.5}
    \begin{tabular}{p{4cm} r r}
        \toprule\\[-5.2ex]
                         Primary type of company & Companies & Participants\\[-0.5ex]
        \midrule
          Enterprise & 5 & 13 \\
 Social media & 3 &13
    \end{tabular}
\end{table}

\begin{table}
   \centering
\caption{The companies that participants worked ranged from relatively small to very large technology companies.}
\label{tab:company_sizes}
  \renewcommand{\arraystretch}{1.5}
    \begin{tabular}{p{6cm} r r} 
        \toprule\\[-5.2ex]
                         Company size by full-time employee count & Companies & Participants\\[-0.5ex]
        \midrule
          Small to medium (300 - 3,000) & 3 & 8\\
 Large (80,000 - 300,000) & 5 & 18
    \end{tabular}
\end{table}

In order to answer our research questions, we conducted qualitative semi-structured interviews with 26 AI ethics professionals based in the U.S. across 8 technology companies. Qualitative interviews allowed us to gain insight into behaviors and practices of individuals, along with their perceptions of themselves and the organizations they are embedded in. Since they were semi-structured, the interviews enabled us to understand standard aspects of AI ethics work across participants, and dig into specific issues most relevant to the experiences of each participant. Given the small size of the field and the context-specific nature of the subject matter, a large-scale survey was not feasible for answering the RQs~\cite{audit_the_auditors}. The first author conducted interviews on Zoom from February to July 2023. Each interview lasted between 30 and 60 minutes.

\subsection{Participant recruiting}
The recruitment criteria and interview protocol were approved by our university's Institutional Review Board. Tables~\ref{tab:participant_roles} and~\ref{tab:participant_employment_status} display the breakdown of participant roles, by role type and employment type. Tables~\ref{tab:company_types} and~\ref{tab:company_sizes} delineate the number of participants and companies by the types and sizes of companies, as of January 2024. We did not link companies with participants in a table in order to reduce the chance of reidentification. We assigned each participant a random numeric identifier, which we include throughout the results and discussion to provide additional context.

Our participant pool comprises a diverse mix of individuals from small to medium and large companies along with social media and enterprise technology companies. They work at companies at different stages of AI ethics implementation. Some have well-defined AI ethics principles, compliance standards, and concrete ethics team workflows. On the other hand, several participants represent their companies’ first efforts in formalizing full-time AI ethics. 

\subsubsection{Inclusion criteria}
We recruited U.S.-based professionals who are currently working or have worked within the last three years on applied AI ethics issues. We view applied AI ethics work as any work that aims to address actual or potential harms from AI products or systems produced or used by a company with external impacts. Thus, we reached out to software and ML engineers, product or project managers, researchers, and those with more specialized functions (e.g. UX design) who worked on AI ethics at least partially in an applied capacity. We did not interview researchers who purely perform research without engaging with product teams or producing material for product and development impact. Some of our participants contributed to research along with other projects. We chose to limit our recruitment as such to narrow in on how ethics work manifests within the product development process. Additionally, we did not target legal or policy teams working on responsible AI or AI policy at companies, as we are interested in the people closest to implementing AI ethics practices for deployed products and services.

We define ``work on AI ethics issues'' broadly to encompass activities such as measuring harms of AI-based products, providing internal guidance on ethics mitigations, managing tooling for mitigations, internal auditing, and red teaming. This range of work takes place under different headings, such as responsible AI, AI ethics, and trust and safety. These varied headings are indicative of a field that has not yet coalesced under a dominant way of working. The trust and safety designation illustrates how AI ethics issues intersect with other issues such as platform health. We note that many trust and safety professionals do not work on AI ethics, and we only recruited those whose job functions centered on AI ethics.

Despite the range of headings and daily activities of these professionals, we group them together due to their dominating shared characteristics and ultimate goal of making AI-based products and services more ethical. Previous research on AI ethics professionals does this as well~\cite{Ali_2023,Rakova_2021,piggybacking,rismani2022plane}. These teams are all relatively new and often work on all the types of activities specified above at different times. They are quite cross-functional and involve similar dynamics engaging with product teams. Altogether, it is sensible to treat these professionals as part of a coherent, though varied, class. 

We recruited professionals who worked full-time, as contractors, or as volunteers on AI ethics issues. Because AI ethics teams are new and small, contractors and volunteer workers play significant roles in shaping AI ethics practices just as full-time workers do. In some cases, they are filling essential roles that their company has not provided enough resources for. Finally, we chose to recruit AI ethics professionals not only who currently work on these issues, but who have in the last three years to reflect the fact that these teams are organizationally precarious and rapid employee turnover is prevalent in the technology industry. In the last three years, there have been recurring layoffs and instances of tumult in AI ethics teams~\cite{msft_layoffs,tech_ethics_layoffs}. By interviewing those in a three-year window, we gain a fuller picture of industry AI ethics. 

\subsubsection{Outreach}
The first author conducted outreach to prospective participants, and recruited through messaging on LinkedIn, personal networks, online technology ethics groups, and snowball sampling with suggestions from interviewees. On LinkedIn, the first author searched for profiles with headings such as ``AI ethics'' and ``responsible AI,'' and locations set to the United States and San Francisco, Seattle, and New York City (as technology hubs). Through participants' LinkedIn profile descriptions of their roles and individual messages with them, we confirmed that participants met the inclusion criteria above.

\subsection{Interview Protocol}
Our interviews were all conducted on Zoom by the first author. Prior to beginning, participants gave their verbal consent to participating in the study as well as their preferences for recordings and level of anonymity. We sought permission to record interviews in order to be more present during the interview and have a source of truth to consult for analysis and quotes. We transcribed these interviews using online privacy-protecting services, and deleted all corresponding video and audio content. Ultimately, we recorded 22 of 26 interviews where we received express permission, and we took notes for the others. We compensated participants with \$25 Amazon.com gift cards after the interview. Regarding privacy, participants could choose their level of anonymity for three descriptors of company, team, and role. For example, all participants gave permission to use their generic role titles, but many wanted to keep their team names or companies entirely private.

To conduct our interviews, we worked off of questions from an IRB-approved interview script developed at the start of the study. The script contains questions about the organizational practices and day-to-day activities of participants and their impacts on products. It can be found in the appendix. At the beginning of an interview, we asked about the participant's team, role, and responsibilities. We then asked questions from the script that were most relevant to the particular activities the participant described they were involved in. Since these were semi-structured, we were able to adapt to the experiences and perspectives offered by participants, and ask further questions to gain insight into specific examples given.

\subsection{Interview Analysis}
To analyze content from our semi-structured interview, we relied on abductive analysis, an adaptation of grounded theory to combine its inductive methods with knowledge garnered from prior relevant theoretical and empirical work \cite{abductive_analysis}. We chose this methodology in order to root qualitative data analysis in existing literature. The first author, who conducted all the interviews, made a first open coding pass to identify descriptive codes. In a second pass, they conducted axial coding, generating a preliminary analysis of how different codes are connected. Next, the first author developed a codebook, with code names and descriptions. We chose to conduct initial coding and codebook development in this way because the first author had the most in-depth knowledge of our participants' experiences and perspectives, having conducted all the interviews~\cite{hci_coder}. The first and second authors discussed the initial codebook and the first author selected two interviews representative of the codebook, which we each coded using the codebook. We refined the codebook after discussion, eliminating redundant codes and sharpening language. We then each coded eight more interviews, which the first author selected to cover a range of codes and participant experiences. We reviewed differences in coding until we felt comfortable having reached consensus about the code interpretations. We decided it would not be meaningful to calculate inter-rater reliability since the conversational coding process itself was the main purpose, leading towards theory-generation, rather than the codes themselves. This choice is rooted in previous literature on qualitative data analysis, which highlights situations in which inter-rater reliability may not be additive~\cite{hci_coder}. The iterative process of discussion helped us better understand our themes and the relationships between them. The first author used the final standardized codebook to re-code the remaining interviews. 

\section{Results}
\label{sec:results}
\begin{table}
   \centering
  \caption{Along organizational and technical dimensions, we detail the challenges and strategies of AI ethics professionals in their path to institutionalization. We also note characteristics of successful states of institutionalization.}
    \renewcommand{\arraystretch}{1.5}
    \begin{tabular}{p{2.75cm} p{5.5cm} p{5.5cm}}
        \toprule
                         & Organizational factors& Technical components\\
        \midrule
          Challenges in \newline institutionalization& \RaggedRight Dependence on informal relationship-building; undefined, opportunistic engagements with product.& \RaggedRight Bridging research and engineering; contextual, one-off tools, metrics, and mitigations.\\
          Strategies for \newline institutionalization& \RaggedRight Prioritizing issues for quick and large impacts; formalizing engagements with product.& \RaggedRight Model documentation and governance; building reusable tools, metrics, and mitigations that are understandable and actionable.\\
          Successful states of \newline institutionalization& \RaggedRight Defined product release procedures; documentation requirements; conceptual frameworks to prioritize work.& Reusable technical components embedded in centralized infrastructure and widely used across an entire platform or multiple products.\\
    \end{tabular}
    \label{tab:institutionalization_results}
\end{table}

In the course of investigating the challenges and strategies involved in instituting and potentially institutionalizing AI ethics practices, we found that AI ethics professionals' limited power leads them to constantly experiment with how they engage with product teams. They test and iterate through strategies quickly as they determine what works. Illustrating this, the majority of our participants, including those on established teams, described their organizational structures and practices as dynamic and responsive [P2, P3, P4, P5, P6, P7, P9, P11, P13, P14, P15, P16, P19, P20, P23, P26]. An ML engineer we interviewed said, “Because the field itself is so new,...you have to come up with these processes and define how these things work… Definitely has to be a little bit more flexible” [P7]. On the technical side, participants frequently described their teams' work as applied R\&D [P1, P4, P5, P7, P9, P11, P12, P13, P19, P20, P22]. As one participant said, “the real responsible AI innovation is happening in products, because product is moving so quickly” [P9].  

At the same time, participants across companies described their ultimate goals to build more scalable and replicable AI ethics practices. Institutionalized AI ethics refers to top-down, standardized, formal processes, company policies, and structures that systematically shape behavior, along with reusable technical components. By investing in “scalable infrastructure” and “automated governance tooling,” AI ethics teams can generate ethics efficiency gains, increasing their potential for impact without requiring additional resources or headcount [P3, P5]. Still, ad-hoc organizational and technical work continue to be relevant as new products present very specific, novel problems. 

\subsection{Challenges in institutionalization - RQ1a }
\begin{quote}
\textit{``We had no actual responsibility for any product. And so every team that was using AI, which is, every team in the company, had to figure out a way to like interact with us, or we had to reach out to them...We had product managers, we had engineers, we had designers. We had all of that stuff, but we didn't have a product. And so what we were supposed to do is go out and find the products to work with.''} [P22]
\end{quote} 
Major challenges in the pursuit of institutionalization include overreliance on informal and interpersonal tactics, precarious collaborations with product, and friction translating AI ethics research into usable tools. 

\subsubsection{Reliance on informal relationship building}
Without formal structures to mandate information-sharing between AI ethics teams and product teams, AI ethics professionals must exert extra effort to keep tabs on product. Our participants discussed the importance of cultivating relationships with product teams to increase visibility and gain knowledge of product initiatives to look into [P2, P3, P4, P5, P7, P8, P9, P10, P11, P14, P15, P16, P20, P22, P23, P26]. AI ethics teams are directly embedded in product or engineering suborganizations, within ML infrastructure or internal tooling groups. With access to more information about product team plans, they can better prioritize their own quarterly goals based on risk evaluation of upcoming product development [P7, P9, P10, P13, P14, P15, P20, P26]. When there are not enough PMs and other leaders to spend time relationship-building, other practitioners such as ML engineers and research scientists perform this work outside of their formal job responsibilities [P2, P4, P7, P8, P10, P14, P15, P22]. An applied researcher who was one of the first hires onto a new AI ethics team described how they spent a lot of their time forging relationships with product. They said, ``We are mostly relying on essentially folk knowledge...to make those connections that, ‘Hey, this is the thing we should audit, and this is a thing that still cannot be audited because of data scarcity issues, because of timeline issues, other sort of alignment issues’'' [P10].

Our participants also emphasized the necessity of building the reputation of their teams by showcasing their deliverables [P4, P5, P6, P7, P8, P9, P10, P11, P15, P16, P22, P23]. One method that several teams tried is developing educational content and workshops to share what AI ethics work looks like. This included presentations explaining different metrics for fairness or analysis of case studies of major public technology ethics failures [P7, P11, P24]. In describing a workshop program, a participant explained that they started pursuing it after deciding, ``Let's just experiment. I'm going to just make this my problem right now, and we're going to see what we can do'' [P23]. They piloted it after spreading the word through interpersonal connections. Interviews revealed that relationships built from informal conversations and educational events led to productive collaborations [P7, P11, P14, P15, P23]. These kinds of experimental initiatives reliant on individual motivation are common as AI ethics professionals know their companies will not prioritize proactive efforts. Although these tactics can be successful, they require AI ethics teams spend significant time away from measurement and mitigation work. 

\subsubsection{Undefined engagement models with product teams}
AI ethics professionals are often forced to accept undefined, ever-changing engagements with product teams due to the lack of accountability structures. Our participants described how their collaborations with product differed project-to-project. An applied scientist at a company with relatively established responsible AI practices mentioned that their team was experimenting with deep-dive embedded engagements with product teams for their last few projects, but that they were each structured entirely differently. “We were brought in at different stages of the product development cycle, given different levels of... visibility into what was going on, and had very different levels of resourcing and prioritization” [P15]. This example reflects the non-systematic way AI ethics resources are deployed and prioritized. 

With no specific regulatory or company policy objectives, those willing to let AI ethics professionals work with their products were often those with personal interests in ethics or fairness, with the time and resources to help answer ethics team questions and provide product access [P2, P7, P8, P11, P14, P17, P22, P23]. One participant noted that when they first joined their team, they spent a lot of time “finding [people] who [would]…let you implement your idea on their products,” and that this often depended on the product team’s priorities and resourcing ability [P2]. This made it so that AI ethics teams could not often independently pursue projects and sometimes lost time interfacing with teams that did not have the bandwidth or interest in collaborating.

\subsubsection{Bridging research and engineering}
One recurring challenge our participants face is translating AI ethics academic and theoretical work to product development. Participants across companies perceived much of the academic literature on issues such as fairness, toxicity, and explainability as inapplicable to their work [P2, P4, P5, P9, P11, P17, P26]. AI ethics team members know that they are working in a “pretty nascent field, and so a lot of the problems that [they are] solving…are pretty novel ones” [P5]. They described how they had exhausted the current state of research on different technical and sociotechnical ethical AI issues from fairness to misinformation. They helped ideate and operationalize new applied research, from methods of stemming unwanted or inaccurate output generation from generative models to new dataset creation for disaggregated measurements [P1, P2, P3, P5, P9, P11, P12, P13, P18, P19, P20, P22, P24, P26]. In a recurring example, as applied researchers and engineers grappled with unavailability or limitations of demographic data, they believed they could not rely on academic fairness literature because much of it assumes access to demographic characteristics [P2, P4, P7, P10, P11, P26]~\cite{survey_nodemog,lahoti2020fairness,neurips_nodemog}. At a social media company facing this issue, the AI ethics team devised a conceptually simple metric to reduce disparity overall between different individual user experiences on the platform that appeared likely to accomplish their end goal of reducing differential demographic experiences [P11]. In a similar case, participants used proxies such as language, region, and zip code to split data into subgroups [P4, P26]. 

Additionally, operationalizing research in product requires a great deal of coordination across research and product since researchers themselves do not necessarily have the skills of liasoning, product strategy, and rapid experimentation in real-world systems [P1, P5, P6, P9, P19, P20, P26]. AI ethics professionals use their discretion and knowledge of how to quickly incorporate technical elements into products to determine how to do this translation. A few participants working on issues related to explainability, interpretability, and accountability described using conceptual and user research about human-AI interaction to put together concrete guidelines for products [P3, P16, P22]. They needed to fluidly move between research and applied product settings, both of which require different skillsets. As such, the applied researchers, engineers, and PMs we interviewed noted that their official job titles may be “researcher” or “engineer,” but in reality the “line between an engineer and researcher was often blurry... a lot of engineers [were] working on research projects [and] researchers [were] writing software tools'' [P11; P2, P4, P9, P26].

\subsubsection{Contextual, ad-hoc tools and measurements}
Because the scope of potential issues is large and resourcing is low, AI ethics professionals may deprioritize developing playbooks or reusable tools to avoid cutting into time addressing urgent problems [P5, P7, P9, P13, P14, P20]. They often must develop highly contextual tools, measurements, and mitigations to apply to specific teams and products' intended use cases. This is especially true in crisis response situations, when AI ethics teams are tasked with monitoring issues during sensitive time periods such as elections or with fixing issues criticized by media [P2, P4, P5, P6, P13, P14]. The tools developed in these cases are not immediately replicable in other systems since they are implemented quickly. Participants emphasized aiming for speed in their engagements by meeting product teams and their products where they are [P6, P8, P9, P13, P14]. Because products often have existing performance measurement pipelines to track product quality, ethics teams might decide to extend those on an ad-hoc basis instead of developing reusable components [P9, P13]. One participant mentioned, “it's a continual challenge to get something that's generalizable enough and sometimes it's better to just build on what teams already have, if it helps them get started sooner” [P9]. 

\subsection{Strategies for institutionalization - RQ1b}
In order to increase adoption of AI ethics tools and processes, gain resources, and promote standardization, AI ethics teams employed a mix of organizational and technical tactics, from creating new accountability structures to improving the usability and reusability of their tools. 

\subsubsection{Prioritization of issues}
Several participants explained how they prioritized the issues they tackled to build support for their team in hopes of gaining resources and working with more teams. Typically, they consider (1) components or products with large surface areas that touch consumers, (2) low hanging fruit in small components, and (3) crises. First, by focusing on issues with the most popular parts of a platform or products with high usage, AI ethics teams can increase the visibility of their teams and prove their impact by citing the large number of people impacted [P5, P7, P11, P13, P14, P15, P20, P26]. As an example, participants targeted high-leverage models, which contribute disproportionately to many downstream decisions and components. This maximizes impact given fixed resources [P7, P14, P26]. Secondly, tackling small or easy-to-fix issues in less used or visible products enabled AI ethics teams to create rapid change [P2, P4, P7, P14]. It also is useful for piloting new techniques or procedures in a low-risk way. One participant described this strategy: ``sometimes there's lower hanging fruit in certain products that people haven't thought as much about, but we can do interesting experiments there and push out ideas more quickly and test them out more quickly'' [P14]. Hackathon projects are one way to initiate these efforts [P2, P17]. Finally, responding well to crisis moments and emergency issues very clearly showed why investment in a strong AI ethics team is essential.

\subsubsection{Formalizing engagements with product}
After recognizing that product teams were not always responsive and collaborative because their goals were disconnected from AI ethics work, our participants described attempts to drive accountability through new structures. They tried assigning point people to product teams, creating consultancy agreements, working agreements, and joint Key Performance Indicators (KPIs), a common goal-setting measurement [P3, P4, P6, P7, P23]. This helped teams establish norms and gain consistent engagement from teams they worked with. Additionally, some participants discussed how they slowly introduced product release processes for product teams. During the first phase, such processes were optional. AI ethics professionals had to go to product teams ``banging on virtual doors...telling people, ‘Hey, you need to come do this thing''' [P9]. Eventually, they became required, with support from leadership. By letting these processes develop over time, AI ethics teams developed best practices and product teams acclimated to new expectations gradually. 

\subsubsection{Model documentation and governance}
Another frequent trend across participants was the importance of model governance as foundational work for future internal audits and interventions. AI ethics teams sought to substantively and systematically keep track of ML models, including major features, datasets, where and how they are used, and who owns them [P4, P7, P9, P10, P12, P16, P18, P22]. Some ethics team members we talked to were tasked with building automated tooling based on this information to help keep it updated with major inputs, outputs, and use cases [P18, P22]. In another example, AI ethics team members worked on surveying model owners and ML engineers as the first step to performing an organization-wide risk assessment [P16]. This type of work is not usually prioritized, so model documentation at technology companies is typically nonexistent or poor. Several participants described how challenging it is to collect this information -- often, there were situations where documentation was incomplete or no one wholly understood the end-to-end of a model because the original developers had left [P4, P7, P16, P18]. By collecting information on models systematically, AI ethics teams could better monitor the ground truth of systems at their companies and identify prominent issues to investigate.

\subsubsection{Reusable tools and metrics}
Participants across the board discussed their desire to increase the impact of their team by creating reusable technical methodologies and tools that could apply to a range of products [P2, P3, P4, P5, P6, P7, P8, P9, P10, P11, P20, P26]. Identifying potential ethics issues and developing metrics for products can be time-consuming, so it is important to scale out measurements when possible. A PM on an AI ethics team reflected this, saying, ``The five-year plan... is to simultaneously invest in scalable infrastructure that lets us duplicate the work that we're going to do in this high-priority set of 20 products to all product teams in the company'' [P5]. We detail the process by which this happens, which we heard echoed from several people. The AI ethics team prototypes a metric or tool, tests it out, and validates it for one product or team. Once it has proven to be useful, it can be packaged up into a reusable form. After having illustrated the value-add and helped other teams implement it for their use cases, AI ethics teams can build the metric into company-wide ML infrastructure or formalize a self-service tool [P5, P6, P7, P10, P15, P19, P20, P26]. By pursuing reusability through an iterative process, AI ethics professionals can ensure their tools are maximally usable and generalizable by gathering feedback at every stage. 

\subsubsection{Aiming for actionability}
Since product teams and engineers are more likely to adopt these new technical tools and metrics if they are easily understandable, many participants aimed to make them simple and interpretable [P4, P7, P8, P14, P15]. One engineer worked on a responsibility tool that provided teams with information on fairness, performance, model explanations, and robustness, among other characteristics. They referenced user studies of the usability of fairness tools to motivate the tool's design. They updated the display of various metrics to be more interpretable to model builders. For example, bar charts might be useful for certain metrics while explicit percentages to drill down into might be better for others [P8].

As another instance of prioritizing interpretability and actionability, an engineer working on fairness measurements discussed how global companies such as theirs have hundreds of possible subgroups they can measure aspects of user experience for. However, it is challenging for a human to quickly look at this number of subgroups and determine what models are better or worse for the desired outcomes. Given this, the team brainstormed higher level metrics to summarize the mass of subgroup data, such as variance of subgroup performance and outliers [P4]. These two examples, echoed by other participants, illustrate how AI ethics professionals are highly responsive to product and engineering needs and attempt to tailor their work for maximum usability.

\subsection{Institutionalized AI ethics - RQ1}
Through our interviews we identified several elements that characterize more mature and institutionally supported AI ethics teams and efforts. These include defined company or organization-wide impact or risk assessment product release procedures, documentation, and reusable, easily accessible technical tools and metrics. 

\subsubsection{Defined release procedures and documentation}
Companies with institutionalized AI ethics practices had developed regimented product release processes for shepherding product teams to deployment and internally auditing their documentation for any ethics issues [P1, P3, P6, P9, P15, P20, P25]. This was especially the case for products flagged as high-risk by legal and policy teams [P3, P9, P15]. There were typically multiple checkpoints of internal review and specific individuals charged with reviewing and approving various stages of product release. One participant talked about how the responsible AI review process was “just another thing that teams have to go through as part of their release process… including going through their privacy reviews, their accessibility reviews, and their security reviews” [P9]. Product teams expected the review, so they had to think about potential harms in advance. As previous work has shown, piggybacking onto existing processes helps normalize and institutionalize new responsible AI practices~\cite{piggybacking}. 

A few AI ethics professionals described how product teams must produce certain documentation when they launch versions of their products to different audiences. Participants on less developed teams either did not discuss documentation as a major part of their workflows or were only beginning to ideate such processes [P7, P10, P16]. Documentation included risk or impact assessments, various performance and fairness metrics, task evaluations, and comparisons to benchmarks. Ultimately, ethics team members and senior leaders used this documentation to form independent interpretations of how products measure up to AI ethics concerns [P1, P3, P6, P9, P13, P20, P24, P25]. Participants mentioned that senior leaders appeared to engage with documentation in detail, flagging possible cherry-picked data or requesting information on criteria for evaluations [P6, P9, P20]. AI ethics team members would evaluate the product documentation and release audience to work with the product team to implement new requirements before current and future releases [P3, P6, P9, P20, P25, P26]. Essential changes would be required immediately before release, but other mitigations that may take longer were assigned lower priority.

\subsubsection{Conceptual frameworks}
In addition to impact and risk assessment documentation, AI ethics teams frequently cited developing conceptual frameworks such as harms taxonomies and risk assessment guidelines [P3, P4, P7, P9, P10, P15, P16, P20, P22, P25, P26]. These are crucial for enabling their teams to scale by prioritizing work more effectively. For example, one participant said of a “governance” workstream of their team, “We were trying to create an internal framework that was efficient, replicable, and at the same time contextual to be able to rank and prioritize our machine learning systems” [P16]. Another echoed the strategy of “[looking] at all the models across [company name] more systematically…a very light high-level ranking of what areas are most concerning” [P4]. These tools helped teams standardize the prioritization process for deciding what projects to work on and how to evaluate the worst possible harms of a system. Instead of reacting to ethics issues when they surfaced as crises or through interpersonal relationships, AI ethics professionals could use systematized tools to triage issues.

\subsubsection{Reusable tools and ML infrastructure}
Once AI ethics professionals developed reusable tools through recognizing common needs in different products, they made them widely accessible by directly embedding them within ML infrastructure or centralized services [P2, P4, P5, P7, P11, P13, P14, P15, P18, P19, P21]. Participants viewed this as an important strategy to achieve broader adoption for AI ethics work across many teams without having to deploy individuals to personally consult with all teams. By implementing new metrics, such as those measuring fairness or toxicity, as part of the standard development process in an A/B testing platform, ethics testing became normalized for engineers. These metrics were automatically included in development and testing workflows. It is much easier to scale ethics efforts when teams running A/B tests can see how business metrics and fairness metrics perform in response to product changes in one place.

Furthermore, reusable and centralized tools, from measurement pipelines to mitigations infrastructure, supports standardization in AI ethics work. One PM on an AI ethics team explained, ``our measurement capabilities are now delivered through a centralized product that vends measurement functionality out to different product teams so that we can basically maintain a single functional source of truth for how we do measurement that we then serve to different product teams instead of having to duplicate the effort over and over'' [P5].  As another example, instead of addressing recurring issues on an ad-hoc basis, AI ethics professionals developed classifiers or blocklists for harmful content and exposed them company-wide via APIs [P13, P15]. Centralizing measurement and mitigation capabilities ensures that issues are treated consistently across a company. Ethics policies already determined by legal and policy arms are more easily enforced.

\subsection{Challenges influencing product decisions - RQ2a}
\begin{table}
   \centering
  \caption{AI ethics professionals face several challenges to influencing product decisions. A few key strategies can help them overcome these, but there are fundamental limitations to their efficacy. }
    \begin{tabular}{p{4.5cm} p{4.5cm} p{4.5cm}}
        \toprule
                         Challenges& Strategies& Limitations\\
        \midrule
          \begin{itemize}[leftmargin=*]
\item Lack of authority of product decision-making.
    \item Communicating the stakes of AI ethics.
    \item Leadership support and \newline company financial wellbeing.
    \item Conflicts over development timelines.
    \item Product, data, and model \newline access and info.
    \item Turf wars.
    \item Pressures to innovate and release novel tech quickly.
\end{itemize}& \begin{itemize}[leftmargin=*]
    \item AI ethics as regulatory \newline response.
    \item AI ethics as product quality assurance.
    \item     AI ethics as crisis mitigation.
\end{itemize}& \begin{itemize}[leftmargin=*]
    \item Issues that do not fit the regulatory response and product quality frames are rarely \newline prioritized.
    \item AI ethics narrows to topics covered by existing or likely regulation, issues that affect large user or product segments, and reactive crisis \newline response.
    \item Using product and user experience as proxies for AI ethics leads to suboptimal decisions.
\end{itemize}\\
    \end{tabular}
    \label{tab:productimpact_results}
\end{table}

Thus far, we have depicted AI ethics professionals as technically and organizationally agile, adapting their approaches to achieve institutionalization. However, their efforts did not always translate into tangible impacts on product. We found that participants could speak a great deal about how they embed with teams, identify issues, and develop measurements. However, when asked about negotiations with product teams and their ultimate decisions, participants often admitted that they did not have much insight into this process, and that ethics considerations are just one component of many that product teams weigh [P2, P5, P7, P8, P11, P13, P14, P15, P17, P21, P23, P25]. Our interviews revealed that AI ethics professionals have much more control over identification and measurement of issues as compared to mitigations. Whereas measurement can lay the foundation for subsequent AI ethics improvements without disrupting any competing product goals, mitigation is more likely to create a conflict, and as a result, is much more difficult to negotiate. 

\subsubsection{Lack of authority over product decision-making}
An applied scientist on an AI ethics team summarized what they were told when they were brought in to help mitigate issues for a major high-priority product launch:  
\begin{quote}
\textit{“You have 3 months before we ship…It doesn’t matter what you find or what you think. Do your best.” } [P15]
\end{quote}
Through our interviews, it quickly became clear that AI ethics professionals often do not have a lot of control over what product teams decide to do as a result of the measurements and analyses provided by ethics team members. As one participant said in reference to a fairness metric that their team embedded into their company’s A/B testing platform, “getting it included is one thing, having people take action on it is another” [P19]. In most cases, AI ethics professionals lack meaningful authority to enforce decisions without real ownership over products. Institutionalized release processes give AI ethics teams more power in negotiations, though there is still ownership friction. Tellingly, the quote above comes from a participant at a company with well-defined AI ethics guidelines and release procedures.

Interestingly, many participants appear to believe that their main value-add is in providing accurate and usable analyses of potential ethics issues to product teams, and that they did not have the background and knowledge to make the ultimate calls on trade-offs and product direction [P2, P4, P5, P7, P8, P11, P13, P15, P21]. In some cases, this self-perception is a result of understanding their lack of product ownership. But several of our participants seemed to endorse their role as information providers, rather than as decision-makers on AI ethics issues. One participant expressed frustration at how product teams compromised ethics to move faster, but said that, at the end of the day, PMs make decisions for product teams and that ``ultimately this is the right way for it to be done. It’s my job to communicate to them how important the responsible AI metrics are, but it is similarly important to have the model actually provide benefits to users” and be cost-efficient [P15].

\subsubsection{Communicating the stakes of AI ethics}
AI ethics professionals face challenges persuasively communicating the harms of AI systems to product teams because risks are often speculative and difficult to concretize or quantify. Product teams are typically focused on possible upside, including factors such as company revenue, reputational benefits, and market share. They look at the business case for building a particular product and spend time fleshing out product feature requirements and timelines in detail. In contrast, risks from AI can be difficult to predict and diffuse. Participants discussed how many product leaders viewed AI ethics concerns as hypothetical and disconnected from concrete, first-order impacts of products [P3, P8, P13, P14, P15, P17, P18, P20, P23, P26]. It can be unclear how to connect potential risks with impacts to the company bottom line. The justifications for AI ethics work are often not powerful since they are ``not the same as saying this can contribute 20\% to our main goal metric by the end of Q3'' [P26]. Additionally, even when a product changes as a result of an ethics team intervention, it is difficult to evaluate the harms that were averted, especially in the case of novel technologies that require more speculation on downstream use cases and risks. This creates an ongoing challenge to maintain funding and support for AI ethics. Ultimately, “until executives truly see the damage,” mitigating potential harms is not prioritized relative to product development [P17].

\subsubsection{Leadership support and company financial well-being}
Whether or not AI ethics teams have power to influence product decisions is dependent on their continued existence. Support from high-level leadership, especially executives, is crucial. Many AI ethics professionals interviewed were painfully aware of the shaky foundations of their teams. One participant noted that it is:
\begin{quote}
    \textit{“clear that these teams…exist under very fragile circumstances. Based off of the whims of people in power and capitalist interests,...[AI ethics teams’] work really can be wiped out overnight.”} [P4] 
\end{quote} 
This is starkly apparent during recessionary periods, when resources are constrained and companies cut from “unnecessary” cost-generating teams. Several of the participants in this study were laid off before this study commenced or even after we interviewed them. They mentioned their worries about the current economic environment and continued support for AI ethics [P2, P4, P7, P10, P11, P16, P19]. One participant noted that a high-level AI ethics full-time employee had been laid off, while they had been informally tapped to continue some of that work with volunteer labor [P17]. 

Beyond existential questions, high-level leadership support for AI ethics and economic conditions also affected AI ethics team internal strategy and mandate. Participants from multiple companies discussed how the personal backing of specific leaders helped propel their team in developing engagements with product by dedicating capital and headcount. AI ethics teams could expand their projects to more teams and take on costly initiatives such as building new datasets [P3, P4, P7, P9, P12, P13, P16, P20, P24, P26]. Additionally, during negotiations with product teams to prioritize features for releases, executives and managers aligned with AI ethics goals help push for ethics mitigations even when they conflict with product leadership desires [P4, P5, P9, P14]. Conversely, increased economic pressure means that AI ethics teams are forced to constrain their ambitions since ethics work is seen as optional [P8, P14, P15, P17, P22]. Product teams that previously were willing to work with ethics teams may stop collaborating due to productization pressure.

\subsubsection{Conflicts over development timelines}
Because product teams have their own goals that are disconnected from AI ethics goals, they do not have clear incentives to incorporate AI ethics work, especially when that would involve specialized efforts and additional time. Product teams are focused on shipping features as fast as possible. Standard agile development at technology companies means that there is flexibility in how developed a product is at each release date, and product teams iterate through to higher levels of functionality. While product may accept bare minimum functionality of an MVP, AI ethics teams attempt to enforce a higher quality standard [P3, P8, P6, P13, P15, P17]. Additionally, product focus is typically on short-term goals, which are scoped to time periods from 2 weeks to 3 months, whereas AI ethics teams want to plan much farther in advance as they anticipate issues and build capacity to deal with them over time [P5, P9, P13, P18, P20, P22, P24]. If ethics initiatives are not formally included in product team goals or if they span long time horizons, they are likely to be deprioritized repeatedly. 

A few participants brought up negotiating over what mitigations are necessary before the launch of different products. They worked with product teams to timeline out goals for mitigations over the course of a few releases, instead of requiring all identified issues be fixed before rollouts [P3, P9, P13, P20]. One participant on an AI ethics team discussed how in the vast majority of cases, product teams seem to think about responsible AI at the very end of the development process. Understanding this, this participant tries to focus on mitigations that can be applied at that point instead of spending too much time recommending more intensive solutions such as collecting new data [P8]. This tension is exacerbated when AI ethics teams lack engineering resources to offer. 

In order to keep product timelines on track, companies may prioritize “band-aid” or short-term mitigations meant to alleviate harms on the consumer-use side without addressing the fundamental technical system and data issues that are root causes [P4, P6, P7, P15, P25]. One instance of this is a company-wide blocklist put in place before public demos of upcoming premier products [P25]. This blocklist broke other systems at the company, indicating that it was not implemented and tested rigorously. Echoing this, another participant said, in reference to a mitigation a product team put into place as an emergency response to a PR issue, “those sorts of band-aid solutions, I think, will be chosen 9 times out of 10 by the product teams when the people who actually work in responsible AI are out of the loop there” [P15]. These mitigations take up less time, allowing product teams to focus on marching forward with new product features. While this can alleviate an issue at surface-level, the fundamental problem remains. Also, the mitigations may not be as replicable in other scenarios or products. 

\subsubsection{Product, data, and model access and information}
One of the consequences of product’s deprioritization of ethics is that AI ethics teams often do not receive the resources they need to generate useful measurements, much less mitigations. Product teams do not have an incentive to go out of their way to explain their systems to ethics team members. In many cases, companies do not prioritize maintaining good documentation, so it can be a challenge to independently learn about the inner workings of certain systems or models without direct engagement with the team owning those. Moreover, there must be usable and accessible methods for testing systems while in development. Sometimes, testing infrastructure is not a priority investment for product teams [P5, P13, P15]. As an example, one participant mentioned that product teams they worked with did not give AI ethics practitioners access to API endpoints that match the production experience. The only way to effectively audit the product was to write custom scripts to scrape data [P15].

\subsubsection{Turf wars} 
When product team goals and AI ethics work conflict, turf wars ensued during the course of negotiations [P3, P5, P12, P14, P15, P17, P22]. In one instance, an applied scientist who had worked on the release of a generative AI product mentioned that their ethics team evaluated possible harms and wanted to restrict particular kinds of sensitive inputs and outputs because of misuse concerns. The product team was not willing to reduce that functionality at all because they saw it as an important, exciting feature to sell. After a back and forth, the product team decided to take the ethics team’s more conservative recommendations for other parts of the system, while keeping the functionality under debate as a compromise [P15]. Ultimately, product makes the final call. In a few instances, AI ethics teams faced challenges to their work when product teams made their own decisions concerning ethics without consultation with the ethics team [P3, P14, P15, P22]. This was caused by organizational dysfunction in some cases, while in others, product teams were frustrated by AI ethics team attempts to ``[tell] us what's right on our platform'' and change their products [P14]. Product teams believed that they had the most expertise on their systems, so they could make the most informed decisions about ethics issues. This led to redundant work or poorly designed mitigations that did not address root causes. Sometimes, AI ethics team members had to completely re-implement or fix these solutions.

\subsubsection{Organizational pressures to innovate and release novel technologies quickly}
AI ethics concerns take a backseat to the primary aims of productizing novel technologies and products as fast as possible, which leads to uneven and reactive investment in AI ethics capacities. This exacerbates many of the above dynamics and makes it even more challenging to build a sustainable, scaled up AI ethics practice. Given that AI ethics evaluations of novel systems may require novel developments or inputs from researchers, research organizations must have corporate support to invest time and resources into building evaluations, new methods, and datasets [P1, P3, P5, P9, P13, P20, P24]. Participants expressed concerns that their companies were moving full-steam ahead to develop AI capabilities and implement them into products without a corresponding investment in ethics work [P1, P17, P25]. A few mentioned that product managers and engineers are racing to incorporate generative AI, whether in-house or through external services, into their products. The desire for rapid deployment impedes AI ethics planning. A PM situated in an ML product organization working on cutting-edge technologies noted that their team tries to stay on top of various product teams’ ship timelines, but that sometimes release dates are kept under wraps [P13]. This makes it difficult to proactively address issues, especially if mitigating them will take a long time and the AI ethics team wants to negotiate over what elements are necessary by a launch.

Notably, a few of the AI ethics professionals we interviewed brought up the infeasibility of mitigating all potential ethics issues in emerging technology products and the “release and review” model of deployment [P3, P9, P13, P20, P21]. Due to constrained resources and the inability to anticipate all potential issues, AI ethics professionals scope down their work to highest priority areas and negotiate with product teams on mitigations. They consequently focus on the most clear and reputationally or legally costly risks [P5, P13, P20, P25]. Several participants working on issues in generative AI systems at different companies noted that certain issues that dominated news post-public release were already known, but were not prioritized to be fixed [P15, P21]. In some cases, this was a result of divergent product team and ethics professional goals. In others, the ongoing conversation on AI ethics centered around the level of risks acceptable and the need to see how people actually use the products instead of making overly cautious decisions [P3, P13, P20, P21]. One participant mentioned that they agreed with their company’s decision to market their new technology as very experimental, because “there’s room for it to mess up and it makes it easier to release.” “Everyone’s aware that it’s going to mess up… there could be some harms, but you don’t necessarily have to shut it down” [P3].

\subsection{Strategies for influencing product decisions - RQ2b}
Given the challenges AI ethics professionals face as they attempt influence product decisions, they lean on three key narrative strategies, transforming ethical concerns into: 
\begin{enumerate}
    \item narrow regulatory problems based on current or impending legislation, 
    \item product quality and user experience issues, and
    \item responses to public relations crises.
\end{enumerate} 

\subsubsection{AI ethics as regulatory response}
Many of the participants we interviewed discussed receiving guidance from legal and policy employees and prioritizing their work based on regulatory risk [P1, P3, P5, P6, P8, P9, P12, P13, P14, P17, P20, P24, P26]. Existing or impending legislation, especially on privacy and civil rights, with specific process-based or outcome-based requirements have led some companies to spin out well-resourced AI ethics or trust and safety workstreams to tackle these. For example, some participants said they have entire dedicated teams to address model documentation to comply with legislation or to mitigate fairness concerns in ad delivery [P5, P12, P18, P22]. Regulatory readiness was the most effective tactic to shift product team activities. As one participant noted, 
\begin{quote}
\textit{“The best way forward is actually to use legal as the arbitrator in order to make some of these things happen, because at the end of the day… if a single lawyer is like,  ‘no, this is a requirement based on regulation,' this will override any amount of executive disgruntlement and disagreement.”} [P17]
\end{quote} 
Frequently, AI ethics professionals would appeal to product teams by noting how ethics work could help the company avoid expensive lawsuits [P5, P8, P12, P17, P18, P22, P24]. Basing AI ethics work on regulation is effective because the stakeholders -- regulators -- and the downside risks -- monetary costs of noncompliance but likely more importantly, reputational damage -- are clear. Speculative risks of AI are concretized. Ignoring these important stakeholders and downside risks would reduce the legitimacy and perceived trustworthiness of a company.

\subsubsection{AI ethics as product quality assurance}
We found that other participants often framed their work through the lens of improving product experience for users. They focused on win-win situations where building more ethical products is also a better business and performance decision, rejecting the bias-accuracy trade-off framing. In some cases, participants explicitly said that they built up the business case as a strategy for achieving product team buy-in [P5, P13, P14, P22, P24]. They wanted to obscure normative ethical questions by using the language and metrics of product in order to avoid contestation. But across companies we also heard participants wholeheartedly endorse equating AI ethics with consistent individual user experience or notions of accuracy (does the model or product do what it is intended to do?) [P3, P4, P6, P8, P10, P11, P15, P26]. They linked ethical AI directly with product success -- for instance, several PMs saw the lack of investment in AI ethics as a blocker to AI innovation, since there was no safe and usable way to deploy models without the coinciding ethics work [P3, P6, P9]. Issues like hallucination and toxicity in language models are thus important to address so that more businesses and consumers choose to use their products [P6, P9, P17]. They see responsibility, both compliance-based and beyond, as a value that consumers care about and recognize when they use these companies’ products. 

A few AI ethics professionals at social media companies described their efforts to promote equitable user experience across their platforms utilizing solutions that “still keep people on the platform or ideally actually improve their engagement” [P11, P5, P10, P14]. Bringing up the hypothetical example of racial bias in creator monetization on their platform, one AI ethics PM saw this as a “big trustbuster for the customer experience,” which would threaten the reputation and growth of the product [P5]. By tackling issues that are aligned with product team goals, AI ethics teams can expect to more successfully persuade product teams to work with them. They can also better prove their value to leadership. If ethics interventions positively impact product goals, there is a business case for keeping the ethics team around.

\subsubsection{AI ethics as crisis mitigation}
During moments of crisis response, such as when an ethics issue has become a reputational risk or is clearly severely harming many people or a platform's health, AI ethics teams are given more control over product decisions than usual. They can take advantage of these moments to apply and generalize their interventions to more products or larger surface areas. They are asked to act fast and are given leeway to implement whatever they deem necessary in the moment [P4, P5, P6, P7, P12, P13, P14, P15, P26]. A few of our participants gave examples of crises where they ``[could] be more heavy-handed in how we fix the model'' [P6]. One engineer working on information integrity issues explained that product teams were more amenable to working with their team during high-risk periods for information flows, such as during times of public unrest or elections [P14]. The public stakes and PR risks are higher during these times, which means companies are relatively more concerned with downside risk and risk mitigation. This engineer also explained how a mitigation they developed to distinguish different types of sensitive content during high-risk periods led into a broader project on sensitive content more generally. Though the increased degree of power in crisis situations is temporary, interventions are stickier -- a reactive workstream can be transformed into a proactive opportunity. 

\subsection{Limitations in influencing product} 
\label{sec:limitations_for_influence}
The frames of regulatory response and product quality are helpful for slotting ethics work more easily in a company’s usual workflows and set of priorities. They align with corporate risk management and general product goals, and thus allow AI ethics teams to skirt around unresolved normative questions. However, these narratives are severely limited when ethics teams consider important issues that fall outside of these buckets. Without a crisis response moment that happens to touch on these other issues, there is little promise of motivating action [P5, P6, P8, P12, P14, P15, P22, P26]. One PM who primarily works on fairness mitigations very clearly laid out this unfortunate reality: 
\begin{quote}
\textit{``if there’s no immediate or medium-term legal risk, and if there’s a strong negative trade-off between fairness and product outcomes, there's just virtually no way to get that work on the road today.”} [P5]
\end{quote}
Though AI ethics professionals may have an ideal set of priorities based on their risk assessments and values, ultimately we found that they make compromises or proactively narrow their requests and projects in anticipation of product team responses.

We heard how AI ethics teams’ use of legal consequences to motivate AI ethics progress narrowed the scope of their work. A few participants expressed frustration that a significant portion of their team goals had moved to legal compliance, leaving other AI ethics efforts on issues such as explainability and user agency under-resourced [P5, P18, P22]. One participant worked on transparency and explainability issues before their team shifted to compliance and mentioned that the change hurt the ability of the team to take a broader perspective on ethics concerns. A primary focus on regulators meant that “We are doing the right thing for the company for the business and for compliance, and not for the people who are actually going to be using these products in the end” [P22].

Because the product quality narrative often hinges on the quantity of users impacted, AI ethics initiatives that target user segments or product surfaces that are relatively small are difficult to justify [P12, P13, P14, P22, P24, P26]. Leaders may question why time is going into those projects, and ethics professionals themselves may be worried about how their impact and performance will be evaluated. As an example, an engineer working on information integrity described a project they wanted to work on to promote better local community content for civic engagement. The product team that owned the product was skeptical and resistant because this “little sliver of content” would not “move their top line goals” [P14]. 

Furthermore, a singular focus on product and user experience can sometimes lead to suboptimal decisions on social responsibility. Individual user experience can certainly be an important indicator of the overall value and performance of an AI system. However, some elements of ethical AI, especially those involving socially contested normative assumptions, combined with the compliance framing of AI ethics, can lead to decisions that reduce direct risk for the company but are socially harmful. For instance, one participant brought up the social harms of erasure that result from decisions to manually block or change outputs based on certain identity terms [P15]. In another example, product leaders and executives wanted to blanket reduce civic and political content on the platform to better entertain users. An integrity team stepped in to caution against this and to draw a distinction between the positive social value of certain civic, political, and news content on the platform [P14]. 

Ultimately, participants were well aware of the limitations presented by organizational priorities of product and regulatory risk management. They expressed hope that regulation and industry standards for AI ethics could help fill this gap.

\section{Discussion}
\label{sec:discussion}
Our results highlight how AI ethics professionals function as values-driven institutional entrepreneurs who strategically attempt to scale their work and gain influence over product. They are highly experimental and agile, adopting tactics as necessary to introduce and institutionalize new organizational and technical practices. However, regardless of how developed these practices are, there are significant blockers in translating them into product impact. At the end of the day, AI ethics teams are resource-constrained and lack ownership of revenue-generating products. Thus they often do not have meaningful authority over product decision-making. Given this limited mandate, industry AI ethics is stuck in a state of minimum viable ethics, wherein the interventions that succeed are narrow, must appeal to clear stakeholders like regulators and product consumers, or appear immediately necessary to avoid PR crises. A broader vision of AI ethics necessitates change in this corporate calculus.

In this section, we first return to literature on organizational change in UX and privacy practices to contextualize our findings on the role of AI ethics professionals. We discuss high-level themes of institutionalization and then offer recommendations for researchers, industry professionals, and policymakers who seek to advance the goals of AI ethics. 

\subsection{Learning from organizational change in UX and privacy practices}
Based on what we have uncovered in this study, the challenges and strategies involved in working on industry AI ethics are similar to the dynamics revealed in analogous privacy and UX research (Section~\ref{sec:rel:othertechpractices}). When a new class of professionals attempts to infuse values into existing structures of a company, it is an uphill battle, especially when those values do not coincide directly with profit incentives. It is useful to specify how these challenges and strategies manifest with the particular set of values of AI ethics. Since the dynamic itself is not unexpected, we can apply lessons here from how the fields of privacy and UX have developed in industry.

First, to emphasize the crucial if uneven role of AI ethics professionals' discretionary power over the values embedded in company structures, we look to the idea of soft resistance. Analyzing UX professionals, Wong argues that ``soft resistance'' is an analytical lens that helps us better understand the double bind that ethics professionals face while working on change from within companies~\cite{softresistance_ux}. While external actors may find the internal strategies as a whole to be lackluster, ethics professionals are viewed as radical within companies. They are constrained to partial resistance as they deploy business-friendly narratives to further their values. Still, ultimately, these ``tactics of soft resistance are about creating organizational change that flows upward from everyday actions''~\cite{softresistance_ux}. Our study reveals how AI ethics professionals can sometimes wield enough power to institute new values-driven practices and even see resulting product changes. They are important levers for change in technology development. Despite the partial and slow nature of their strategies, organizational change from within can succeed, especially if complemented with external efforts.

Secondly, literature on the institutionalization of privacy practices forewarns us of the threat of corporate capture of AI ethics work, both internally and externally. The concept of legal endogeneity describes how law can emerge from organizational practices within the area it aims to regulate. Edelman originally conceptualized this in the context of antidiscrimination law~\cite{edelman_antidiscrimination}. In the absence of specific requirements and definitions related to discrimination in employment, corporate interpretation of the law led to ceremonial structures and practice that undermined antidiscrimination law. In a recent book, Waldman shows how this applies to privacy law, which has been fundamentally shaped by technology companies. This is due to ``legislative ambiguity and process-oriented rules,'' which ``give corporate professionals – lawyers, consultants, and compliance experts, for example – the chance to define what the law means and protect their employers''~\cite{ari_ground}. In this case, individuals within companies are the first to determine what compliance means and how their organizations can achieve that state. For instance, Waldman discusses how GDPR mandates that collecting or processing data is unlawful unless companies justify this with at least one of six possible reasons. One is the company's ``legitimate interests.'' There is very little description of what this entails, leaving industry professionals with significant leeway to argue for a definition that suits their corporate interests.

Building on the framework of legal endogeneity in the context of privacy law, Waldman describes how corporate responses to regulation then further shape our legal regime. The legal interpretation of privacy laws develops downstream of specific cases of noncompliance brought to courts to adjudicate. In the absence of specific requirements, regulators increasingly take the existence of symbolic structures and ceremonial practices as evidence for compliance. In courts, judges may move from simply referring to these organizational elements in decisions to describing their existence as relevant in determining compliance. Finally, they may defer to structures as compliance~\cite{ari_ground}. For instance, the mere presence of privacy policies on websites or standardized processes of data use review indicates superficial compliance but nothing about outcomes. 

We apply these two relevant insights from UX and privacy literature to our findings. We acknowledge the essential role of empowered industry AI ethics professionals in advancing values-based organizational change, but note potential pitfalls of overreliance on internal efforts without robust regulation. From our results and developments in technology policy, it is clear that AI policy is subject to legal endogeneity. AI ethics professionals are crucial in shaping and propelling particular strategies for dealing with harms of AI. Given this, we first discuss overarching insights on institutionalization from our findings along with recommendations for effective corporate AI ethics work. Finally, we turn to policy implications.

\subsection{Recommendations and insights to strengthen organizational practices}
Below we provide recommendations and insights that can help advance AI ethics work in industry. Many of these are distilled from the strategies we heard from study participants. A few are relevant for researchers who aim to boost applied AI ethics with their work. The majority should be useful for AI ethics professionals themselves as well as industry leaders who seek to empower these teams.

\subsubsection{Generalizable practices are easier to institutionalize, but contextual, one-off experimentation is still essential} 
Both technical and organizational practices can be more readily scaled if they are widely applicable to a range of products and teams across a company. Companies with centralized infrastructure and shared components will more easily institutionalize technical practices. Organizational practices like red-teaming that can be adapted to many different domains are similarly easy to spread. However, only relying on generalizable practices leaves gaps in AI ethics coverage. Potential harms, threat models, technical implementations, and deployment strategies differ widely depending on the products at hand. For this reason, contextual, specialized technical work is still necessary to investigate issues. This is especially the case with new technologies or use cases. On the organizational front, it can be challenging to develop playbooks for AI ethics for emerging technologies. Experimentation as issues emerge is crucial for developing best practices over time.

\subsubsection{Leadership buy-in is more important for institutionalizing organizational practices than technical practices}
Leadership support is necessary when AI ethics teams introduce new organizational practices. These practices often require collaboration and alignment on goals across many different functions and teams. This is a logistically difficult task that must be incentivized by top-down backing. New practices like product release procedures can typically only become standardized and effective if mandated from above. On the other hand, if product teams are amenable, AI ethics teams can independently perform technical work since they are usually set up as technical teams and have the expertise needed. There is a lot of latitude for low-level efforts to reach out to product teams and pitch ideas. Coordinating between two teams is much less arduous than negotiating with several layers of leadership. Furthermore, because the technical components built by AI ethics teams are likely compatible with existing technical infrastructure and engineering processes, scaling them is not a unique or overly disruptive ask of technical leadership. 

\subsubsection{More research should emphasize applied settings}
Our interviews revealed an urgent need for research that can more easily transfer to industry settings. For instance, common issues that would benefit from applicable research include dealing with missing, partial, or approximate data, the lack of or inability to use demographic data, and ways to improve models in post-deployment scenarios. Broadening coverage of AI ethics research will boost AI ethics professionals' abilities to deal with real-world problems with well-evidenced methods. Transferable research also requires clear communication on reproducing methods and delineating what scenarios they might apply in. 

\subsubsection{Improve the usability of tools and metrics} 
One key recurring concern throughout the study was how usable, and thus actionable, AI ethics tools are. End-users should be able to easily interpret and use them. Within a company, consultations with product teams as a particular tool or metric is built would be helpful. In general, it should be clear how tools and metrics can integrate seamlessly with existing products or development systems. In the case of evaluations tools, design aspects such as the prioritization scheme for how different issues are surfaced are important. For example, the number of users impacted or severity of harm can inform this.

\subsubsection{Formalize collaborations with product teams through concrete goals} 
In order to strengthen the incentives of product teams to work on AI ethics issues, ethics professionals should create formal cross-team goals embedded in the official goal-setting framework product teams use. Standardized structures for accountability are essential for AI ethics work to be prioritized, rather than left as informal, precarious projects. A lot of time can be wasted when AI ethics professionals engage with product teams whose collaboration is dependent on goodwill.

\subsubsection{AI ethics teams must have dedicated engineering resources}
AI ethics work is propelled when engineers are formally allocated to ethics teams. With dedicated engineers, these teams can take control of their projects from timelines for completion to implementation details. It is easier to convince product teams to collaborate because AI ethics teams can be more independent. There is also more bandwidth to address issues across many teams.

\subsubsection{Monitor outcomes of AI ethics interventions}
As we have seen, it is not enough to successfully implement an AI ethics ship review process or new educational initiative without tracking resulting changes. We highly encourage AI ethics teams to monitor the outcomes of their work through time and try to hold their companies accountable for this. For example, we can measure how well a particular bias metric integrated in an A/B testing framework changes product team behavior by tracking the frequency it surpasses a threshold and the triggering update is still approved.

\subsubsection{Invest in model documentation and governance initiatives}
A major blocker to AI ethics work is weak or nonexistent model documentation. Without a birds-eye view of models across a company along with detailed documentation about their creation, usage, and issues, AI ethics teams struggle to assess possible risks, let alone determine how to address them. Technology companies should invest in efforts to document models better. Long-term benefits of this include easier and more systematic AI ethics work along with reduced redundancy in engineering across the many teams.

\subsubsection{Break down AI ethics initiatives into short-term goals} 
As we learned, accomplishing AI ethics progress can necessitate long incubation periods, in comparison to the short timelines of product feature development. There are several reasons to break AI ethics goals into smaller chunks. First, this will enable steady progress towards these goals in a way that aligns with the pace of product team development. Additionally, because experimentation is a key element of industry AI ethics, piloting new practices in small ways, such as low visibility product segments, is useful to test and iterate on them. Finally, even short-term or partial efforts can widen buy-in for future change and improve products.

\subsection{Policy implications}
Emerging developments in AI policy are likely to strengthen corporate incentives to act on and institutionalize AI ethics. However, regulatory frameworks that hinge on voluntary compliance will face the same fundamental limitations detailed in this study. Recent AI policy in the U.S. and the E.U. detail high-level principles and expectations of AI systems, alongside recommendations for operationalizing these~\cite{voluntary_commitments,blueprint,nist_rmf,eu_ai_act,execorder}. These emphasize many practices that are aligned with aspirational, institutionalized AI ethics. This includes defined documentation requirements, continuous monitoring and evaluation systems, and the proactive testing of products pre-launch. Companies that are compelled to follow through with these initiatives will invest resources in developing reusable technical systems and governance procedures. This will help AI ethics teams reduce time spent in the ad-hoc, disorganized phase of building out their work. Additionally, regulation will increase the decision-making power of AI ethics teams since concrete standards and consequences for AI ethics will incentivize product teams to prioritize those issues. 

Nonetheless, most current approaches to policy fall into similar traps as above, in which procedures and measurements are emphasized, leaving it to company discretion to implement mitigations and respond based on what is surfaced. This work will rely on AI ethics professionals' abilities to have impact in their organizations. As we have seen, this is highly precarious. Although policies based on voluntary compliance may shift norms in the institutional environment to create an expectation that technology companies develop responsibly, relying on this lever perpetuates the status quo of ethics based on goodwill and reactivity. Furthermore, AI policy is subject to legal endogeneity, paralleling the development of the privacy legal regime. Regulation that is overly vague will very likely lead to deference to standards proffered by industry. 

Based on our findings, we offer a few suggestions for policy initiatives. First, along with procedural requirements, more specific standards for accuracy, fairness, and explainability should be developed, especially for highly sensitive domains. Instead of relying on companies to determine appropriate standards, policymakers should collaborate with academics, civil society, and industry groups to specify these in legislation. Next, policy should establish meaningful consequences for noncompliance. As several of our participants stated, legal mandates to comply are the most effective way of intervening in the product development process. Data privacy and civil rights policy applied to technology illustrate how enforcement through financial and legal liability can transform ``nice to have'' practices into ``must haves''~\cite{fb_propublica}. Along these lines, as we have seen, established policy on data privacy, civil rights, consumer protection, and antitrust can be used to advance AI ethics work. Regulators and legislators can pursue this strategy, equipping AI ethics professionals to have greater product impact.

Additionally, smart AI policy initiatives can bolster internal AI ethics teams by strengthening the external counterparts to this work such as academic researchers and independent auditors. Policymakers should direct more resources to applied research problems in general or to specific research problems relevant to industry settings, as specified in the above recommendations section. Furthermore, cultivating an empowered third-party auditing and open-ended research ecosystem can both improve the ability of AI ethics teams to succeed internally as well as unlock potential fruitful partnerships between these different actors. Researchers and third-party auditors can perform similar work as industry AI ethics teams if they have sufficient access to the data, models, and systems at hand, and they can develop tools and methods that are more likely to transition into widespread practice. Legislation that specifies access expectations for researchers for certain high-risk systems can support this. Mandating better tools such as sandbox environments that enable system probing with safeguards will naturally boost AI ethics teams' abilities to test. Growing the auditing ecosystem overall will produce many downstream benefits.

\subsection{Future work}
There is much more to uncover as industry AI ethics develops over time. We highlight a few areas that warrant further exploration that were only touched on by our study. 

\subsubsection{Perspectives on risk, ethics, and emerging technologies}
Our participants frequently mentioned the need to balance risks and benefits of novel technologies. It is challenging to do this given the uncertainties involved in innovation and the diffusion of new products. Release guidelines have emerged in recent years to help manage this~\cite{Solaiman_2023,anthropic_responsiblescaling,openai_frontierrisk}. Analysis of these along with case studies will be useful in iterating on these further. Additionally, future interview-based and conceptual research should flesh out what anticipatory ethics entails in real-world scenarios~\cite{nanayakkara2020anticipatory,anticipatoryethics}. ``Anticipation work'' involves practices that imagine potential futures and guide us towards them, handling uncertainty~\cite{anticipationwork}. In applying this to industry AI ethics, we can map how uncertainty is managed within organizations hoping to reduce, but not completely eliminate, risk in their work.

\subsubsection{Trust and safety in comparison to AI ethics}
The relationship between trust and safety and AI ethics in industry is underexamined. Our interviews showed how, within organizations, professionals with responsibility in these areas often overlap in mandate and perform similar work. Yet, these areas have quite different cultures and approaches. Trust and safety teams are common in the domains of social media and highly consumer-centric technologies where users can create or exacerbate problems, while AI ethics generally encompasses a broader category of concerns but specific to AI applications. It would be useful to disentangle these two similar areas of work to identify differing assumptions, threat models, or approaches that change outcomes. Additionally, studies can point out redundancies in work and potentials for collaboration.

\subsubsection{Affective experiences of AI ethics professionals}
Our results reflect the focus of our interview questions -- organizational structures, practices, and outcomes of specific team endeavors. Future research can deeply investigate the individual perspectives and emotional experiences at play in AI ethics work. Many of the strategies AI ethics professionals use in their work rely heavily on aggressive personal initiative and emotional labor. Due to all the blockers to achieving impact, this work may be exhausting and demoralizing. Studies that dive into the personal relationships these professionals have towards their work, as well as how they manage the emotional experience with their careers will help us better understand and strengthen these teams. Also, learning more about the perspectives they bring, which have been informed by their previous life and work experiences, can shed more light on how they approach their work.

\subsubsection{Quantitative and causal research}
In addition to descriptive studies on industry AI ethics, research that systematically tries to draw out causal connections between organizational practices, governance models, and mitigations and ethics outcomes will be useful. This is especially the case for evaluating the efficacy of regulation. Quantitative studies based on large-scale surveys of AI ethics practitioners will help identify major issues. Finally, impact assessments of specific types of mitigations such as certain harm detection tools are also promising. 

\section{Conclusion}
As AI development and productization speed ahead, it is crucial that AI ethics matures at a similar pace. In this study, we focus on how AI ethics and governance professionals, who have limited power within technology companies, navigate their constrained mandate to gain visibility, recognition, and influence. They opportunistically experiment with organizational and technical practices to help them effectively collaborate with product and ultimately aim for institutionalization. Despite successes of scale, we still find stubborn gaps in translating this to product impact due to differing incentives amongst product and AI ethics teams. Effective AI regulation -- guidelines paired with concrete enforcement capabilities -- has the potential to move us beyond a state of minimum viable ethics to robustly safe, just, and trustworthy AI development and deployment.

\begin{acks}
We are grateful for the support of the Princeton Center for Information Technology Policy in developing this project. We also thank Emily Cantrell, Klaudia Jaźwińska, Varun Rao, Jakob M{\"o}kander, and the members of the Princeton Bias in AI discussion group for helpful feedback on drafts of this paper and study design.
\end{acks}

\bibliographystyle{ACM-Reference-Format}
\bibliography{references}

\appendix
\section{Interview Instrument}
Below is the guideline script for all interviews. We ultimately decided to skip the first section on backgrounds and theories of change of AI ethics professionals since we wanted to focus our limited time on the other two sections on ways of working and impacts. As these were semi-structured interviews, we did not ask all of these questions, but rather dug into specific subjects and questions based on what each participant brought up. 

\textbf{Backgrounds and theories of change of AI ethics professionals}
\begin{itemize}
    \item Can you tell me a bit about your education and professional history?
    \item What ethical, social, technical issues or scenarios are you most worried about when you think about AI in general and in the context of your company?
    \item Are there particular texts, media, or people (e.g. other professionals within this company or others, civil society leaders, policymakers, academics, etc.) that influence how you think about AI governance and ethics? How so?
    \item Do you have a strategy / theory of change for impacting AI product development and deployment at this company? What is it?
\end{itemize}

Any other follow-up questions in this category would follow the above themes of understanding how the interviewee thinks about AI harms and strategies for dealing with them within a company.

\textbf{Where and how do AI ethics teams and professionals operate within a company's existing product development process?}
\begin{itemize}
    \item What does your current team work on? What is your specific role and set of responsibilities? 
    \item What are the major goals of your team? (set by leadership; internal sense of goals)
    \item Where do/does you/your team sit in relation to the rest of the company? (reporting structure, what organization or sub-team this team sits in, how your team relates to product/engineering teams)
    \item Can you describe the process for how you/your team tackle a new project or task? (e.g. do you actively find product teams to work with or do teams come to you?)
    \item Is there any requirement for teams to engage you or your team in ethics-related work? If so, in what situations does it apply?
\item If applicable, what percentage of your/your team’s work is applied in response to issues found in production versus to issues found in the development or pre-deployment phases?
\item How is your work impacted by potential or actual news about the company or similar organizations? (Does crisis response play a role?)
\item At what stage of development or deployment do you usually engage with product/engineering teams? 
\item What is your level of access into the products that you/your team work with? (inputs, outputs, system design, internal models, UI, product roadmaps, etc.)
\item Are there particular tools or frameworks you/your team uses in this process? If yes, can you describe them? (e.g. checklists for compliance, red teaming methods, auditing tools)
\item How do you use them to identify actual or potential harms or issues?
\item Are these tools/frameworks used across the entire company for all products? (Can ask for specific tool/framework mentioned)
\item What are the major challenges you and/or your team face in accomplishing your goals? (e.g. organizational buy-in, access to data, tools, resources, anticipating harms, etc.)
\end{itemize}

Any other follow-up questions in this category would follow the above themes, and would dig deeper into specific metrics, tools/frameworks used, organizational dynamics, and processes used.

If the participant does not formally work on AI ethics as per their official role responsibilities but does this work informally/on the side or implements interventions that come from AI ethics professionals, the above section will be amended to reflect that and to ask about how this work fits into the interviewee’s formal role responsibilities, how it is prioritized and evaluated, and how they engage with formal AI ethics professionals.

\textbf{Impacts}
\begin{itemize}
    \item Can you describe a specific case/project you/your team took on and how you tackled it, step-by-step?
    \item How is your/your team’s impact measured and evaluated? 
    \item Where have the solutions/interventions/actions of you/your team been applied in the product lifecycle? (e.g. software/system changes, input data changes, deployment changes, etc.). 
    \item How often do you/your team provide input that results in a meaningful product change?
    \item Who or what teams make the final decisions around how you/your team’s interventions are applied to a given product or process?
    \item Has you/your team’s previous work impacted:
    \begin{itemize}
        \item future product versions or different products? (with more or different functionality or with more powerful or different underlying AI systems/capabilities)
    \end{itemize}
    \begin{itemize}
        \item the entire company? (its decision-making systems, formal frameworks, development processes, types of products/projects undertaken, etc.)
    \end{itemize}
\end{itemize}

Any other follow-up questions in this category would follow the above themes and dig deeper into specific cases the participant has described.

\end{document}